\documentclass[aps,prl,twocolumn,showpacs]{revtex4} 

\usepackage{graphicx}
\usepackage{hyperref}
\usepackage{amsmath}

\begin{document}

\title{Statistics of random lasing modes and amplified spontaneous emission spikes in weakly scattering systems}

\author{X. Wu and H. Cao}

\affiliation{Department of Physics and Astronomy, Northwestern University, Evanston, Illinois, 60208\\}

\begin{abstract}
We have measured the spectral correlations and intensity statistics of random lasing modes in weakly scattering systems, and compared them to those of the amplified spontaneous emission spikes.  Their dramatic differences revealed the distinct physical mechanisms. We find that local excitation of a weakly scattering system may greatly reduce the number of lasing modes even without absorption outside the pumped region. The lasing modes can be very different from the quasimodes of the passive system due to selective amplification of the feedback from the scatterers within the local gain region.   
\end{abstract} 

\pacs{42.55.Zz,42.25.Dd}
\maketitle

Random laser has attracted much interest over the past few years. One important subject is the nature of random lasing modes, which was not fully understood. For random systems in or close to the localization regime, the lasing modes are the quasimodes of the passive system \cite{vanneste_2001,genack_2005}. The mode formation rely on the interference of scattered waves which return to the same coherence volume via different {\it closed} paths. Thus scattering provides coherent and resonant feedback for lasing, leading to the selection of lasing frequencies \cite{caoPRL99}. Surprisingly, random lasing with resonant feedback was also realized in diffusive or even ballistic systems, despite the coherent interference effect was expected to be negligibly small \cite{frolov_1999,ling_2001,polson_2005,wu_2006}. Although the lasing modes were assumed to be the quasimodes with small decay rates \cite{burin_2001,apalkov_2002,patra_2003,hacken_2005,lagendijk_2006}, recent theoretical studies suggested that the quasimodes of a passive random system may not be the genuine normal modes of the same system with gain \cite{deych_2005,tureci_2006}. Experimentally local pumping is necessary to observe discrete lasing peaks, which is hard to explain if the lasing modes were the quasimodes spatially extended across the entire system. Although they could be the anomalously localized modes \cite{apalkov_2002}, the lasing modes are not as rare as expected. Our latest model of the absorption-induced confinement of lasing modes in the pumped region \cite{yamilov_2005} does not apply when the absorption outside the pumped region is negligible. 

In addition to the lasing peaks, sharp random spikes were observed in the single-shot spectra of amplified spontaneous emission (ASE) from colloidal solutions over a wide range of scattering strength \cite{mujumdar_2004}. The spikes were stochastic and varied from shot to shot. They were attributed to single spontaneous emission events which happened to take long {\it open} paths inside the amplifying random medium and picked up large gain.  Thus the emergence of ASE spikes did not rely on resonant feedback or coherent interference. However, a clear distinction between the stochastic ASE spikes and coherent random laser modes is still missing.  

In this Letter, we present a systematic experimental study to illustrate the difference between the ASE spikes and lasing peaks. While the ASE spikes can appear in the absence of scattering, the lasing peaks require the coherent feedback provided by scattering. The ensemble-averaged spectral correlation functions for the ASE spikes and lasing peaks, as well as the statistical distributions of their spectral spacing and intensity, are shown to be very different. Such differences underline their distinct physical mechanisms. We find that local pumping of a weakly scattering system may greatly reduce the number of lasing modes even without absorption outside the pumped region. This increases the frequency spacing of lasing modes and facilitates the observation of discrete peaks in spectrum. The lasing modes, though still extended across the entire system, can differ dramatically from the quasimodes due to selective amplification of the feedback from the scatterers within the local gain region.  
 
Our experiments were performed on the diethylene glycol solutions of stilbene 420 dye and TiO$_2$ particles (mean radius = 200nm). The experimental setup was the same as that in Ref. [8]. The stilbene 420 was intentionally chosen for the weak reabsorption of emitted light outside the pumped region. The absorption length $l_a$ at the center emission wavelength $\lambda_e$ = 427nm was 6cm at the dye concentration $M$ = 8.5mM. It was much larger than the dimension ($\sim$ 1cm) of the cuvette that held the solution. At the particle density $\rho = 3\times10^{9}$cm$^{-3}$, the scattering mean free path $l_s \simeq 1.3$mm at the pump wavelength $\lambda_p$ = 355nm, and $l_s \simeq 1.0$mm at $\lambda_e$. Although $l_{a} \simeq 10\mu$m at $\lambda_p$, the pump light penetrated much deeper than $l_a$ due to the saturation of absorption by intense pumping. The excitation volume had a cone shape of length a few hundred micron and base diameter 30$\mu$m. Because the cone length was smaller than $l_s$, the excitation cone in the colloidal solution was almost identical to that in the neat dye solution. For the emitted light, the transport was diffusive in the colloidal solution whose dimension was much larger than $l_s$. Light amplification, however, occurred only in a sub-mean-free-path region. The motion of particles in the solution provided different random configuration for each pump pulse, which facilitated the ensemble measurement under identical conditions. 
   
\begin{figure}
\centerline{{\scalebox{0.8}{\includegraphics{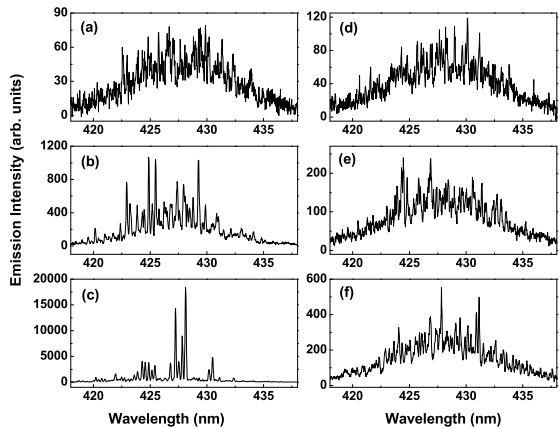}}}}
\caption{Single-shot emission spectrum from the 8.5 mM stilbene 420 dye solutions with $3\times10^{9}$ cm$^{-3}$ TiO$_2$ particles (a)-(c) and without particles (d)-(f). The pump pulse energy is 0.05 $\mu$J for (a) \& (d), 0.09 $\mu$J for (b) \& (e), 0.13 $\mu$J for (c) \& (f).}
\label{fig1}
\end{figure}
The single-shot emission spectra from the colloidal solution are shown in Figs. 1(a)-(c) with increasing pump pulse energy $E_p$. At $E_p$ = 0.05$\mu$J [Fig. 1(a)], the spectrum exhibited sharp spikes on top of a broad ASE band.  From shot to shot the spikes changed completely. The typical linewidth of the spikes was about 0.07nm. The neighboring spikes often overlap partially. As pumping increased, the spikes grew in intensity. When $E_p$ exceeded a threshold, a different type of peaks emerged in the emission spectrum [Fig. 1(b)]. They grew much more rapidly with pumping than the spikes, and dominated the emission spectrum at $E_p = 0.13\mu$J [Fig. 1(c)]. The peaks, with  the typical width of 0.13nm, were notably broader than the spikes. Unlike the spikes, the spectral spacing of adjacent peaks was more or less regular. We repeated the experiment with solutions of different $\rho$ as well as the neat dye solution of the same $M$. The peaks could only be observed with particles in the solution, while the spikes appeared also in the spectrum of emission from the neat dye solution [Fig. 1(d) - (f)]. Although they were similar at $E_p = 0.05\mu$J, the emission spectra with and without particles  were dramatically different at $E_p = 0.13\mu$J. Even under intense pumping, the emission spectrum of the neat dye solution had only spikes but no peaks [Fig. 1(f)]. The maximum spike intensity was about 50 times lower than the maximum peak intensity in the colloidal solution at the same pumping [Fig. 1(c)]. While the pump threshold for the appearance of peaks depended on $\rho$, the threshold for the emergence of spikes in solutions with low $\rho$ was similar to that with $\rho = 0$. 

In our previous experimental and numerical studies \cite{wu_2006}, we concluded that the large peaks represented the lasing modes formed by distributed feedback in the colloidal solution. Although the feedback was weak at low $\rho$, the intense pumping strongly amplified the backscattered light and greatly enhanced the feedback. In contrast, the feedback from the particles was not necessary for the spikes which also existed in the neat dye solution.  Thus the spikes were attributed to the amplified spontaneous emission. 
 
\begin{figure}
\centerline{
\begin{tabular}{c c}
\includegraphics[width=0.235\textwidth]{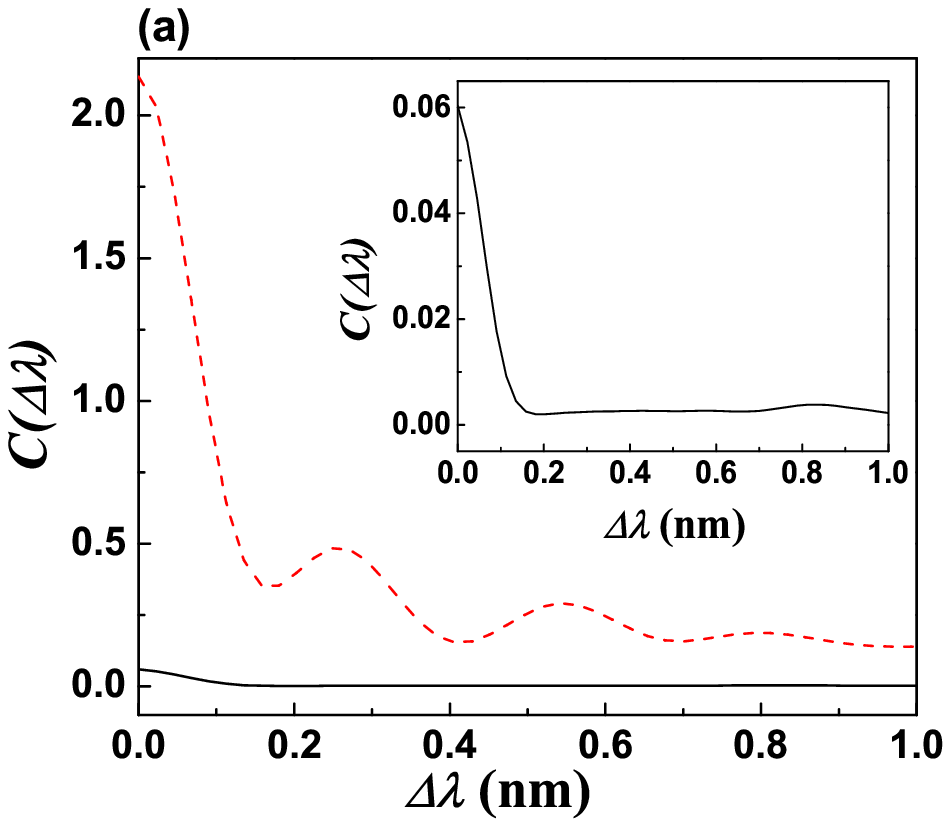} & \includegraphics[width=0.235\textwidth]{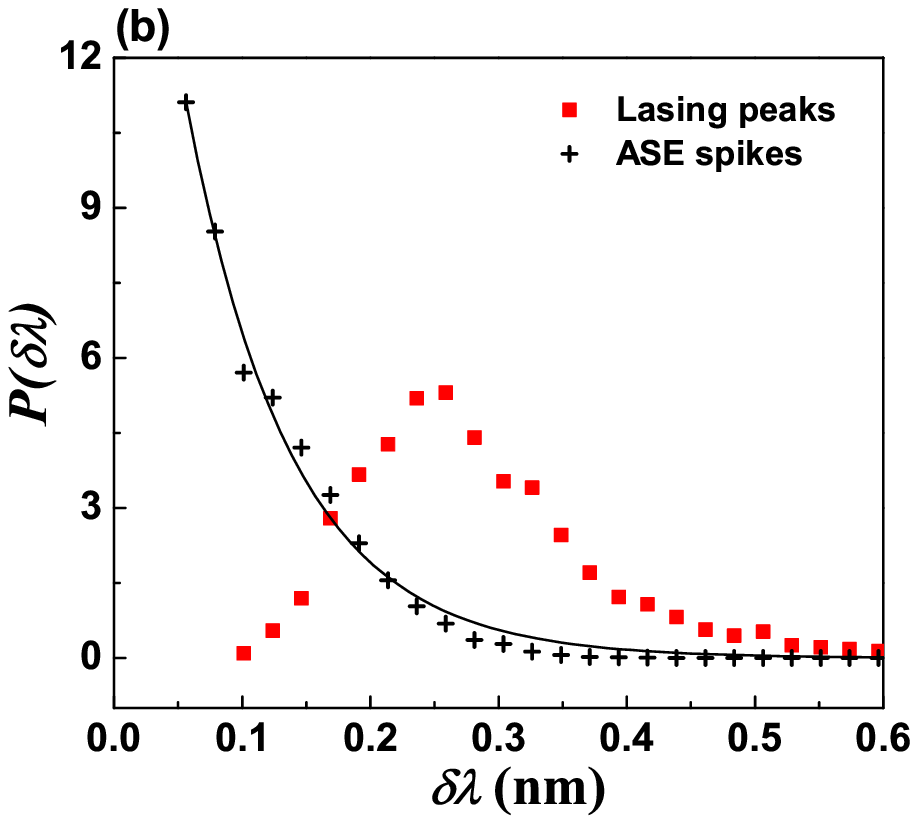}
\end{tabular}}
\caption{(a) Ensemble-averaged spectral correlation function of single-shot emission spectra for 8.5 mM stilbene 420 dye solutions with $3 \times 10^{9}$ cm$^{-3}$ TiO$_2$ particles (dashed curve) and without particles (solid curve). The  pump pulse energy is 0.13 $\mu$J. (b) Statistical distribution of wavelength spacing between adjacent lasing peaks (square) and ASE spikes (cross) obtained from the same spectra as in (a). The solid curve represents the exponential fitting $P(\delta \lambda) \sim \exp(-\delta \lambda / 0.082)$.}
\label{fig2}
\end{figure}
To demonstrate quantitatively the differences between the ASE spikes and lasing peaks, we investigated their spectral correlations and intensity statistics. Since it was difficult to obtain reliable statistical data for the ASE spikes from the colloidal solution at high pumping due to the presence of dominant lasing peaks, the data of ASE spikes were taken from the neat dye solution instead. We checked that at low pumping where the lasing peaks had not appeared, the statistical data for ASE spikes collected from the colloidal solution with low $\rho$ were similar to those from the neat dye solution. The ensemble-averaged spectral correlation function $C(\Delta\lambda)$ was obtained from 200 single-shot emission spectra acquired under identical condition. We chose the wavelength range 425-431 nm, within which the gain coefficient did not change much, to compute $C(\Delta\lambda)= \langle I(\lambda)I(\lambda+\Delta\lambda) \rangle / \langle I(\lambda)\rangle \langle I(\lambda+\Delta\lambda) \rangle -1$. At low pumping [Figs. 1(a) \& (d)] $C(\Delta \lambda)$ for the neat dye solution was very similar to that for the colloidal solution with low $\rho$, while at high pumping [Figs. 1(c) \& (f)]  $C(\Delta \lambda)$ became significantly different as shown in Figure 2(a). $C(\Delta \lambda = 0)$, which was equal to the intensity variance, had a much larger value for the colloidal solution than for the neat dye solution. This result reflected the intensity ratio of the lasing peaks to the background emission was much higher than that of the ASE spikes. Moreover, $C(\Delta \lambda)$ for the colloidal solution exhibited regular oscillations with the period $\sim$ 0.27nm. Due to slight variation of lasing peak spacing, the oscillation was damped and the correlation peaks were broadened with increasing $\Delta\lambda$. Nevertheless, the oscillation of $C(\Delta \lambda)$ survived the ensemble average despite the lasing peaks changed from shot to shot. This result confirmed not only the lasing peaks in a single-shot  spectrum were more or less regularly spaced, but also the average peak spacing was nearly the same for different shots. In contrast, $C(\Delta \lambda)$ for the neat dye solution was smooth and decayed quickly to 0 as $\Delta \lambda$ increased from 0 [inset of Fig. 2(a)]. The ASE spikes produced irregular oscillations in the spectral correlation function of a single shot emission spectrum. However, such oscillations were removed after average over many shots. This result reflected the stochastic nature of the ASE spikes.

We also obtained the statistical distribution $P(\delta \lambda)$ of wavelength spacing $\delta \lambda$ of the ASE spikes and that of the lasing peaks. Fig. 2(b) represented $P(\delta \lambda)$ extracted from the same data as Fig. 2(a). Due to the finite width of spikes/peaks, we could not get $P(\delta \lambda)$ at $\delta \lambda$ close to 0. $P(\delta \lambda)$ for the ASE spikes could be fitted approximately by an exponential decay. It suggested the ASE spikes satisfied the Poisson statistics, which meant the frequencies of individual ASE spikes were uncorrelated. Instead of an exponential decay, $P(\delta \lambda)$ for the lasing peaks reached the maximum at $\delta \lambda \simeq 0.25$nm. This result reflected the spectral repulsion of lasing modes. Such repulsion could not be explained simply by mode competition for gain \cite{cao_2003}, because the average mode spacing depended on the length of excitation cone. Although the particle suspension was in the diffusive regime, $P(\delta \lambda)$ for the lasing modes did not fit the Wigner-Dyson distribution \cite{Beenakker_1997}.
 
\begin{figure}
\centerline{
\begin{tabular}{c c}
\includegraphics[width=0.235\textwidth]{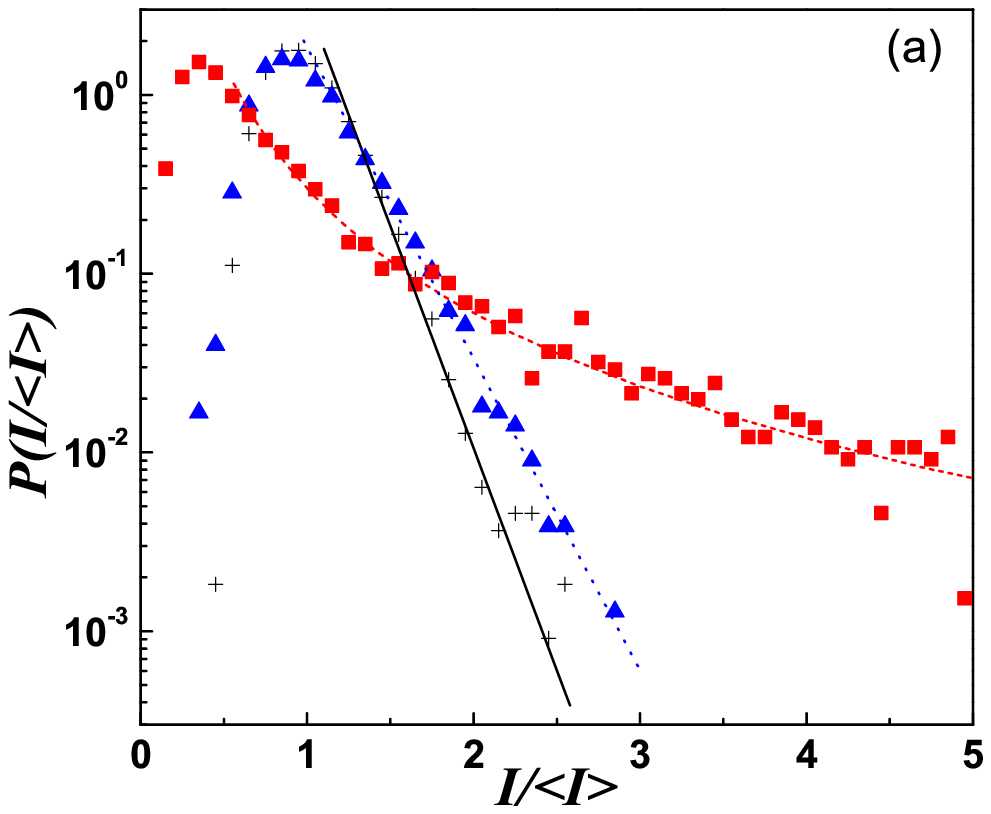} & \includegraphics[width=0.235\textwidth]{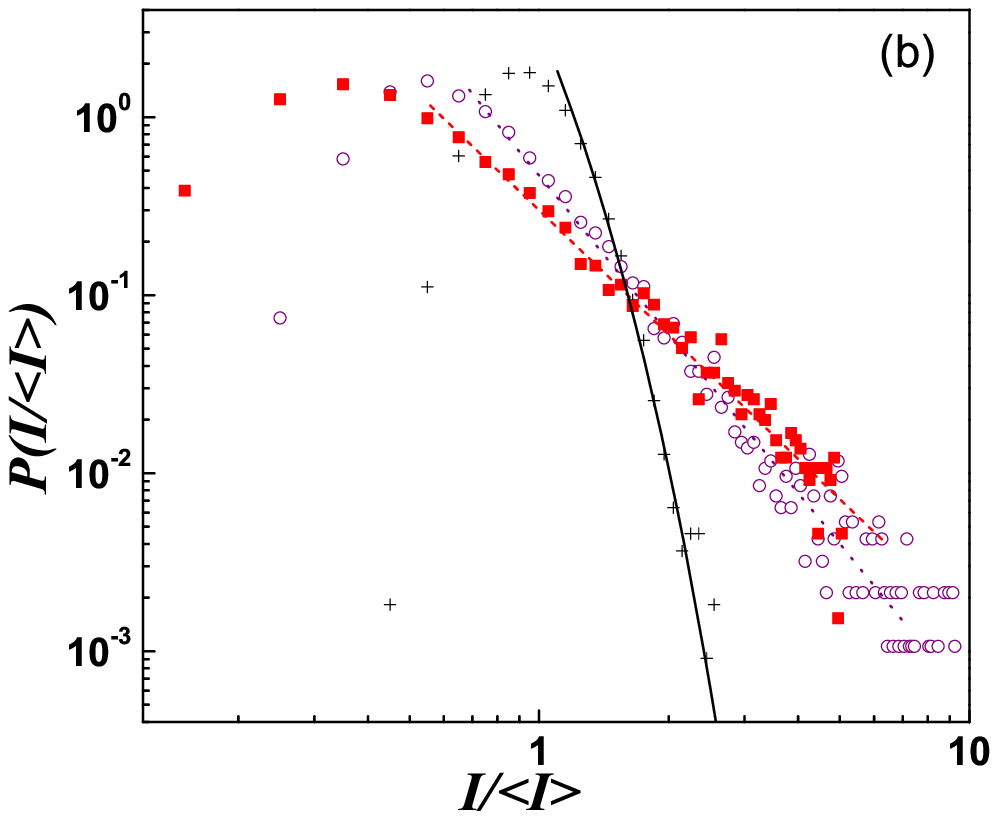}
\end{tabular}}
\caption{(Color online) Statistical distributions of the intensities of lasing peaks (square, circle) and ASE spikes (cross, triangle), obtained from the 8.5 mM stilbene 420 dye solutions with $3\times10^{9}$ cm$^{-3}$ TiO$_2$ particles and without particles, respectively. The pump pulse energy is 0.09 $\mu$J (circle), 0.13$\mu$J (square, cross), 0.39 $\mu$J (triangle). The lines represent fitting (see the text).}
\label{fig3}
\end{figure}
Further difference between the ASE spikes and lasing peaks was revealed in the statistical distributions of the intensities ($I$) of lasing peaks and ASE spikes. The log-linear plot in Fig. 3(a) clearly showed that $P(I/ \langle I \rangle)$ for the ASE spikes had an exponential tail at large $I$. The tail became more extended as $E_p$ increased. The solid and dotted curves in Fig. 3(a) represented the exponential fitting $P(I/ \langle I \rangle) \sim \exp(-a I/ \langle I \rangle)$ at large $I$ for $E_p = 0.13\mu$J (cross) and 0.39$\mu$J (triangle) with $a = 5.7$ and 4.0, respectively.  The log-log plot in Fig. 3(b) revealed that $P(I/ \langle I \rangle)$ for the lasing peaks had a power-law decay at large $I$, $P(I/ \langle I \rangle) \sim (I/ \langle I \rangle)^{-b}$. The fitting of the data at $E_p = 0.09\mu$J (open circle) and 0.13$\mu$J (solid square) gave $b = 3.0$ and 2.3, respectively. Hence, the power-law decay became slower with increasing pumping. 

The above experimental results demonstrated the fundamental difference between the ASE spikes and lasing peaks. Next we presented a qualitative explanation for some of the data. The stochastic structures of the pulsed ASE spectra of neat dye solutions were observed long ago \cite{sperber_1988}. In our experimental configuration, the ASE spikes originated from the photons spontaneously emitted by the stilbene molecules near the end of the excitation cone at the beginning of the pump pulse. As they propagated along the cone, these photons experienced the largest amplification due to their longest path length inside the gain volume. The ASE at the frequencies of these photons was the strongest, leading to the spikes in the emission spectrum. Although the spontaneous emission time was a few nanosecond, the 25ps pump pulse created the transient gain and only the initial part of the spontaneous emission pulse was strongly amplified. Thus the ASE pulse was a few tens of picosecond long, followed by a spontaneous emission tail. The spectral width of the ASE spikes was determined by the ASE pulse duration. We extracted the average width of ASE spikes from $C(\delta \lambda)$ in Fig. 2(a). After taking into account the spectral resolution of our spectrometer, we estimated the ASE pulse duration to be 10-100 ps, in agreement with the pumping duration. Since different ASE spikes originated from different spontaneous emission events which were independent of each other, their frequencies were uncorrelated. It led to the Poisson statistics of the frequency spacing of neighboring ASE spikes. Although the occurrence of ASE spikes did not rely on scattering, multiple scattering could elongate the path lengths of spontaneously emitted photons inside the gain volume and increase the amplitudes of some spikes. In our experiment the size of gain volume is less than $l_s$, the effect of scattering on the ASE spikes was negligibly small, thus the ASE spikes exhibited little dependence on the particle density.  

To interpret the data for the lasing peaks, we must understand how the lasing modes were formed with local pumping and negligible reabsorption. We calculated the quasimodes of 1D random systems and the lasing modes under global or local pumping using a method described in Ref.\cite{wu_2007}. This method was valid for linear gain up to the lasing threshold, with both gain saturation and mode competition for gain being neglected. Thus the number of lasing modes $N_l$ we calculated represented the maximum number of lasing modes that a random system could possibly have. If optical gain existed only in part of a weakly scattering system and there was no absorption elsewhere, $N_l$ was less than the number of quasimodes of the passive system. This was very different from the case of uniform gain where the lasing modes had one-to-one correspondence with the quasimodes. Moreover, local pumping made the spatial profile of a lasing mode very different from that of the quasimode. Although the lasing mode was still extended across the entire random system, the relative weight of its intensity distribution in the part far from the gain region was reduced. Such differences could be understood as follows. Because the feedback from the scatterers within the pumped region was selectively amplified, its contribution to the lasing mode formation was greatly enhanced. Meanwhile, the feedback from the scatterers outside the pumped region was not amplified, and its contribution to lasing became relatively weak. 

The aforementioned calculation results provided qualitative explanation for the experimental data. Tight focusing of the pump light greatly reduced the number of potential lasing modes $N_l$. The increase of mode spacing facilitated the observation of discrete lasing peaks. Since the reduction of $N_l$ was a result of local amplification, it did not rely on the reabsorption of emission outside the pumped region. The decrease of $N_l$ with the size of local gain region agreed qualitatively with the experimental observation of wider lasing peak spacing for shorter excitation cone (with larger $M$). Since the reabsorption outside the pumped region was extremely weak, the feedback from the particles in the unpumped part of the solution was not negligible. Hence, the lasing modes were not equal to the quasimodes of the reduced system defined by the gain volume. Our numerical calculation also showed that the frequency spacing of adjacent lasing modes under local pumping became more regular as the scattering strength decreased. Such regularity was intrinsic and not caused by mode competition for gain. Because of the dramatic difference between the lasing modes and the quasimodes, $P(\delta \lambda)$ for the lasing peaks [Fig. 2(b)] did not fit the Wigner-Dyson distribution, which held for the quasimodes of the colloidal solution without dye. The statistical distribution of decay rate of the quasimodes could not be used for the calculation of $P(I/ \langle I \rangle)$. Moreover, the mode competition and gain saturation, as well as the initial spontaneous emission into individual modes, must be taken into account. The rapid variation of gain in time and space make the explanation of intensity statistics even more difficult not only for the lasing peaks but also for the ASE spikes. We hope our data will stimulate further theoretical work.  

In summary, we demonstrate experimentally the spectral correlation and intensity statistics for the lasing modes in a weakly scattering system are very different from those for the ASE spikes. Since the ASE spikes originate from independent spontaneous emission events, their frequencies are uncorrelated, leading to Poisson statistics for their spectral spacing.  The lasing peaks represent the lasing modes, which can be drastically different from the quasimodes due to local pumping. Even without reabsorption outside the pumped region, the number of potential lasing modes may be greatly reduced by local excitation of a weakly scattering system. The selective amplification of the feedback from the scatterers within the gain region can make the spatial profile of the lasing mode very different from that of the quasimode. 

The authors acknowledge Profs. A. Yamilov and A. Chabanov for stimulating discussions. This work was supported by the National Science Foundation under Grant Nos. DMR-0093949 and ECS-0601249.

\end{document}